# Analyzing Cascade Sizes of Stopped Projects in SourceForge: Is SOC Theory Applicable to OSSOCs?

Jianmei Yang[1*], Xiangdong Pan[1], An Zeng[2], Hua Bai[1], Bill McKelvey[3], and Zengru Di[2]

[1] School of Business Administration, South China University of Technology, 510640 Guangzhou, Guangdong, China

[2] School of Systems Science, Beijing Normal University, 100875 Beijing, China

[3] UCLA Anderson School of Management

*Corresponding author: fbajyang@scut.edu.cn

**Abstract**

Based on three rounds of data extraction, we first construct complex network models to identify the cascade sizes of stopped projects and their distributions in SourceForge (March 2000 to February 2013). We then analyze and discover characteristics of these cascade sizes and their distributions: most cascade sizes are 1; two extreme sizes coexist; and the cascade sizes in each Period of SourceForge's peak phase exhibit a power-law distribution while lacking scale-free properties. Finally, we discuss the limitations of this study and their implications for Self-Organized Criticality theory.

**1. Introduction**

About 40 years ago, most businesses were conducted in buildings and their employees and formed real communities; however, many modern businesses operate totally on the Internet, such as open-source software (OSS), Wikipedia, and others. The virtual internet communities formed by the people involved in these businesses are called online communities (OCs).

Open-source software online community (OSSOC) is one of the most famous OCs of knowledge production and learning. An OSSOC is a website that allows its users to collaborate with each other in developing software and share their open-source products with others (Maggioni 2002). It is a system consisting of people and projects constructed on software-developing digital platforms.

Founded in 1991, the Linux is the earliest established OSSOC. Given the development processes of Linux, Raymond (1999) presents the "cathedral mode" and "bazaar mode" of software production. An OSSOC is an emerging bazaar-mode software-production organization based on OCs (Lonchamp 2005; Maggioni 2002; Ratkiewicz et al. 2010), which can emerge only in the era of the Internet (Arazy 2016;

Greengard 2015; McQuivey 2013; Tapscott 2015).

The SourceForge OSSOC (abbreviated as SourceForge; sourceforge.net) is one of the earliest OSS hosting platforms. It offers services such as mailing lists, discussion forums, and download servers, allowing users to create OSS projects for organizing and coordinating their software development activities. SourceForge is an important milestone in OSS history, dating back to 1999, and was once almost the only choice for hosting and collaborating on OSS projects; it peaked around 2010; and in the 2020s, its influence has declined compared to modern alternatives like GitHub and GitLab, but it still is an important OSSOC.

As an OSSOC, SourceForge differs from hierarchical organizations in that it operates as a self-organizing evolutionary bazaar. The cascade sizes of its stopped projects (cascade sizes, for short) serve as important indicators of the OSSOC's evolutionary state. Therefore, this paper takes the cascade sizes and their distributions as the subject of study based on our early work (Pan et al., 2015). The stopped projects in this context refer to projects for which there is no longer any activity in SourceForge. Why they are called "stopped projects" and the criterion for judging them are described in Section 3.

Self -Organized Criticality (SOC) is one of the most famous complexity-theories. This theory studies the evolution of a Bak-style self-organized system (BSOS, see Section 2) by using the method of distribution of avalanches (sizes and durations). Bak's "sandpile", a foundational example of SOC, demonstrates the process that a BSOS self-organizes itself into a critical state with sand-avalanches of all sizes (Bak 1996, pp. 49). The critical state is the attractor—the state it is efficiently trying to achieve (Bak 1996, pp. 58, 126-127, 198). An attractor exhibits fractal characteristics when demonstrating scale-free properties; such attractors typically form through the "principle of least effort" (defined as the most efficient behaviors) (Zipf 1949; Bak 1996, p.177). A BSOS must possess some key elements of the self-organization process (see Section 2). The SOC theory has successfully explained a number of avalanche-like scale-free phenomena in nature and society, such as earthquakes, species extinction, forest fires, stock market crashes, and traffic jams (Bak et al., 1993; Bak, 1996, pp. 182, 192, 197-198).

SourceForge can be metaphorically described as Bak's sandpile, where new projects (new sand grains) continuously enter and old projects (old sand grains) cascade out. At first glance, it appears to possess the key elements of a BSOS (see Section 2). Therefore, identifying its cascade size distributions should lead to an investigation of whether it has reached or is evolving toward SOC. However, we do not do so. This is because we question the applicability of SOC theory to OSSOCs (see Section 6). Due to this skepticism, this paper uses "cascade" sizes instead of "avalanche" sizes to avoid misunderstanding.

Research on complex networks has developed over 25 years (e.g., Watts and Strogatz 1998, Barabási and Albert 1999). Complex network models are widely used in many fields. In this paper, we use complex network models to identify the cascade sizes in SourceForge (see Section 4).

The remainder of this paper is organized as follows: First, relevant concepts and methods are presented. Then, based on three rounds of data extraction, cascade sizes in SourceForge and their distributions in each Period (see Section 4) are identified and analyzed. Finally, the study's limitations and issues related to SOC theory are discussed.

## 2. Theory

**SourceForge metaphors**

SourceForge can be metaphorized as a "sandpile" of projects (the people involved are embodied in the projects), in which new projects are equivalent to adding new sand grains to a sandpile. These equivalents to new sand grains bring in additional kinds of energy. SourceForge also exists in Bak's ecosystem. Like species in ecosystems (Bak 1996, p. 133), projects in SourceForge can survive only when their minimum requirements are met; each project has its own minimum requirements of technology and developers; projects with the least fitness of technology and fewest developers will be the first ones to be stopped.

**Definition of BSOS**

Following SOC theory, we define a BSOS (Bak 1996, p. 31) as a dynamical system having the following key elements of the self-organization process: (1) it should be a large dynamical system with many components, the components interact primarily through short-range forces rather than global coupling (Bak 1996, pp. 5, 37). This is because myriad successive individual adaptation events can drive a system of individuals into the SOC state (Bak 1996, p. 137). The critical state is a cooperative phenomenon where the different components of large systems act together in some concerted way (Bak 1996, p. 37); (2) the system must be driven by an external input (energy) (Bak 1996, pp. 5, 37, 49) at a rate much slower than its relaxation time (Bak 1996, pp. 2, 50,177 ); (3) energy must be continuously dissipated to prevent runaway effects (Bak 1996, pp. 5, 49, 51; Prigogine 1955); and (4) the system's evolution is determined through internal dynamics, requiring no external fine-tuning (Bak 1996, pp. 31, 48, 169) .

Bak's sandpile model (Bak 1996, p. 5) is a well-performing example of a BSOS. A BSOS can reach a critical state and exhibit characteristic SOC behaviors such as power-law distributions, scale-free avalanches, and fractal structures. But many BSOSs may not have reached their critical states, as pointed out by Bak (1996, p. 31).

**SourceForge seems to possess the key elements of a BSOS**

First, the number of projects in SourceForge is enormous. For instance, when we conducted the second round of data extraction on February 28, 2014 (Table 1), approximately 419,300 projects were listed on the website. Projects join the community for collaboration; therefore, technological and/or people's links exist between many projects (i.e., short-range interactions rather than global coupling). This indicates that SourceForge possesses the first key element of a BSOS. Second, projects or people (such as project owners, developers, contributors, users, and community members) can join the community relatively slowly and freely, bringing new energy to the system. This means the system is driven by external input at a rate potentially much slower than its relaxation time. The system also dissipates energy, such as when a project fails. These observations demonstrate that SourceForge satisfies the second and third key elements of a BSOS. Third, from the perspective of participants (people), SourceForge functions as an online "bazaar" where participant bazaar behavior is unaffected by any external factors, thus the participants are self-organizing. Since participants operate the projects, project behavior resembles participant behavior, and therefore the projects are also self-organizing. This shows that SourceForge exhibits the fourth key element of a BSOS.

**Concept of SourceForge Cascades**

Different projects in SourceForge are often connected via technology and/or common principal developers. Therefore, after a project is stopped, the fitness of its connected projects will be reduced, causing these projects eventually also to be stopped. This stopping process repeatedly happens among many projects, resulting in SourceForgs cascades similar to the small to large avalanches of sand-grain movements down the edge of a sandpile.

According to the nature of the actual stopped-project process in SourceForge, the size of a cascade is determined by the interactions of technology and developers of stopped projects. A cascade lasts from the first stopped project that causes other projects to be stopped to a project that no longer causes any other project to be stopped. The size of a cascade is defined as the number of projects that are stopped during this cascade. Note that if a project being stopped does not cause any other project to be stopped, it is defined as a cascade with size one (i.e. one-project cascade). As time goes on, along with the addition of new projects and the cascades of stopped projects, the average slope of the metaphoric sandpile of projects becomes steeper. If the average slope eventually reaches a certain angle and cannot increase any further, the BSOS has reached its critical state (Bak 1996, p. 51). If the BSOS has not yet reached its critical state, Bak predicted that it is still in an evolutionary process towards its critical state.

## 3. Data description and defining stopped projects

Data description: We conducted three rounds of data extraction from the SourceForge website (see Table 1). The study of cascade size characteristics is based on the datasets obtained from the first and second rounds. The dataset extracted in the third round aids in analyzing the findings derived from the previous rounds. We gathered data for all types of projects from March 2000 to July 5, 2008 (as the number of projects was relatively low during this period) and specifically for stopped projects from July 6, 2008 to February 2013.

**Table 1. Three Rounds of Data Extraction**

| Round | Data extraction time | Last update time of projects | Number of collected stopped projects | Number of stopped projects | Number of all projects |
|---|---|---|---|---|---|
| 1 | 2009/7/5 | 2000/3 to 2008/7/5 | 28,703 | 52,276 | 152,402 |
| 2 | 2014/2/28 | 2008/7/6 to 2013/2/28 | 114,016 | About 200,000 | 419,300 |
| 3 | 2014/5/15 | 2011/11/18 to 2013/5/15 | 194,991 | About 370,000 | 427,150 |

Defining Stopped Projects: Prior to IEEE Std 2675-2021 (2021), no international standards organization had published a formal classification standard for open-source projects based on their lifecycle status. Although CHAOSS, Apache, and CII, among others, have proposed similar concepts—for example, in 2008, Apache established the Apache Attic to house unmaintained projects, demonstrating its existing internal project governance standards for such classification—none of these were expanded into an industry-wide standard. We extracted project data from SourceForge for the period from March 2000 to July 5, 2008 (the last update date of the projects), with the data extraction performed on July 5, 2009. Given the actual SourceForge situation at that time, where most projects were in the initial/experimental stage and lifecycle categories were limited, we defined "at least one year without any updates" as the criterion for determining stopped projects (see below) for our first round of data extraction. It should be noted that, by current standards, our "stopped projects" include three categories defined in IEEE Std 2675-2021: Inactive (>1 year without indications of maintenance), Archived, and Abandoned projects.

The process of proposing the criterion: we define a random sample of the projects in the dataset and study the time-interval between two adjacent updates of a project in its version control system[1] (VCS). We find that most of the time-intervals are relatively short. 81.5% of the updates occur within two months, the longest time-interval is just five months. But in our third round of data extraction, in May 2014, we find that, among the 195,000 projects without any update from May 2013 to May 2014, 538 of them were later updated between May 2014 and August 2014. This is to say, there are at least 538 projects with an update time-interval longer than one year. Based on these findings, we consider at least one year without any update as the criterion of a stopped project in our study (this criterion is based on the SourceForge "current" situation and should be revised if this time-interval is shortened or lengthened). If the project hasn't been updated at least one year before the date of our data extraction, we define it as a "stopped project." Otherwise, it is still an "active project." Furthermore, if a project has released at least one formal file, we consider it as a successful project. So in the stopped projects there are some successful projects.

Every SourceForge project website shows the basic information of the project, including authors, the registration date, the "last update" date, project's description, number of downloads, comments, and whether formal files have been announced and so on. To extract these data from SourceForge, we use a software called LocoySpider (see the provided link in the Reference section) that we used to extract data from March 2000 to February 2013—this is the time-range based on the last-update date on this website. Our data extraction is in compliance with the terms and conditions of the SourceForge website.

## 4. Methods

**Method 1: Selection of the time interval for each Period**

We investigate the evolution of SourceForge by dividing its developmental process from March 2000 through February 2013 (the last update date) into four distinct periods, with its state remaining relatively constant within each period. The earliest record of a stopped project is March 2000. From March 2000 to July 5th, 2008, SourceForge remained in its initial state. There are very few links among the stopped projects, therefore, we label this time duration as Period I. From July, 2008, to February 2013,

---

[1] "Version control system" refers to the software tools that define how revisions can occur and their nature as a computational model progresses through its various stages.

SourceForge gradually developed to its peak, which was divided into Periods II to IV based on the longest duration of a cascade.

We use "snapshots method" to find the longest duration of a cascade: Based on SourceForge data from March 2000 to July 2008, we first select a time-interval and identify the several "snapshots"[2] of it; and then compare one snapshot with the snapshot in the previous time step, if the number of stopped projects is about the same, it means that the cascade has already finished; we finally test different time-intervals by taking the snapshots and find that most time-durations of cascades are within half a year. Furthermore, we confirm this result using the data from our third round of extraction. Based on these analyses, we take half a year as the time interval for each of Periods II to IV.

**Method 2: Using complex network models to identify cascade sizes**

We construct complex network models of stopped projects to indentify the cascade sizes in SourceForge. Since our collected data consist of a large number of stopped projects, how to identify the stopped projects belonging to a particular cascade is difficult. To deal with this problem, we created an approach to construct the complex network models of projects. But, when we try to construct the complex network models of projects to determine the affected projects in each cascade, we first have to define the relation between projects.

*Links between Projects*

SourceForge relies on modular production. Therefore, the projects in SourceForge have strong technology links (e.g., project A depends on project B). Hence, we use a content analysis method to uncover these links. Specifically, we extract the technology links between projects by analyzing the text in a project's "Description." Moreover the development of some projects in SourceForge needs collaboration among people in other projects. One principal developer may participate several projects, hence, the projects are linked when they have common principal developers (including project owner).

---

[2] A "snapshot" quickly offers information about time-intervals and whether a cascade has ended or not.

*Construction of models*

By considering projects as nodes in a network and connecting two projects if they have a technology relationship, we can obtain a complex network model based on technology relations: since the technology relation among projects is asymmetric, the network model is directed. If project A depends on B, a directed link points from node B to node A. (2) We can also connect two projects if they share common principal developers. Since the developer-relation among projects is symmetric, this model is an undirected network.

We combine the above two models to obtain a complex-network model based on mixed relations. In this model, the nodes are still projects. If two projects have at least one kind of relation, there is an edge between them. In this way, the edges in the network can have different meaning. Some represent the technology relations; some are based on common principal developers. For some edges, both relations exist. But whatever relation an edge represents, we can say there is a "*neighbor link*" between the pair of nodes. For simplicity, a network based on mixed relations is defined as an undirected one.

Furthermore, in order to identify cascades sizes, we have to know which the stopped projects are and their "*neighbor links*". There are two methods for finding this information. The first one (1) is to find the "neighbor links" of all types of projects and then "stopped" projects. The second one (2) is to find "stopped" projects and then their "neighbor links." Hence: (1) construct a network model of all types of projects so as to find their neighbors and *then* find the stopped projects among them; or (2) first find stopped projects, and *then* construct the network model among these stopped projects to find their neighbors. In these two methods, the meaning of the nodes in the constructed networks is different. For Period I, we construct the network model of all types of projects, for other Periods we construct the network models of stopped projects.

*Cascade sizes and connected graphs in a complex network*

Figure 1 illustrates the relationship between a cascade and a connected graph in a stopped-project network. In part I of this Figure, project *A* is developed solely by person *a* and also uses project *B*; *B* is developed by persons *b* and *c*; project *C* is developed by persons *c* and *d*; and project *D* is developed only by person *e*. Therefore, *c* is developing two projects at the same time. In this situation, if *c* leaves this community, both project *B* and *C* will be stopped due to the common principal-developer relation. As project *A* relies on project *B*, stopped *B* will lead to *A* stopped. Therefore, a cascade is formed. The cascade includes projects *A, B*, and *C*, so its size is 3. In part II of this Figure, the stopped projects *A, B*, and *C* form a simple network that just has 3 nodes and one connected graph with size 3. Thus, the cascade size is equal to the size of the connected graph in the network.

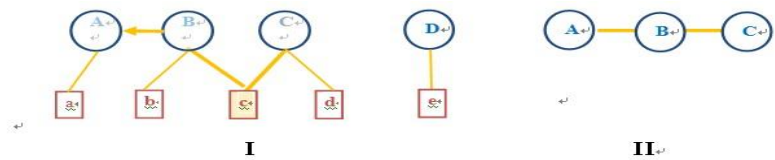

**Figure 1. An Illustration of the relationship between a cascade and a connected graph**

A complex network often consists of many connected graphs (Figure 2). For each of Periods II to IV, we need to find all connected graphs within the corresponding stopped-projects network and then compute the size of each connected graph. In this way, we can identify all cascade sizes (i.e., number of nodes in a connected graph) for each of Periods II to IV.

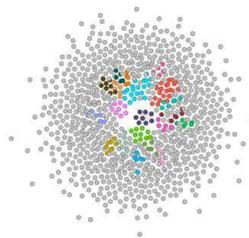

**Figure 2. An Illustration of Connected Graphs in a Complex Network** (The different colors denote connected graphs that have different numbers of linked members, i.e., within each colored circle the members of the project are linked; members in the grey circles are not linked)

**Method 3: Measurement of scale-free characteristics**

We know that a Power Law (PL) distribution does not specifically indicate the scale-free characteristic of a dataset, except when the PL exponent is within a certain range (the absolute value of the exponent is less than 3) (Barabási and Albert 1999, Yang et al. 2013).

We can analyze the scale-free characteristic of a dataset through the following two steps: first, measure the heterogeneity of the dataset; then, for datasets with heterogeneity, measure the span of the dataset described by a PL distribution (SOPLD). The concept of heterogeneity indicates that individuals in a system can possess very different quantities of a specific attribute, i.e., these quantities can differ by several orders of magnitude (OOMs). The more OOMs contained in the relevant dataset, the greater the heterogeneity of the system from the perspective of that attribute. We define the number of OOMs contained in a dataset as the system heterogeneity indicator of relevant attribute. The SOPLD can be represented by an indicator that is the ratio of the number of OOMs contained in the data fitted by a PL to the number of OOMs contained in the entire dataset with heterogeneity. The closer this ratio is to 1, the stronger its scale-free characteristic is.

By synthesizing the above two aspects, we can deduce that a system's scale-free characteristic can be evaluated by the absolute value of the PL exponent. If it is less than 3, we may conclude that this system has a scale-free characteristic, the smaller the absolute value, the stronger is the system's scale-free characteristic. Also, the absolute value of the PL exponent has a lower limit.

**5. Results and findings**

First, Table 2 and Figure 3 show the properties of the SourceForge network models in each Period. The properties of the stopped-project network models are the foundation for analyzing the cascade size characteristics. Although the Table and Figure only contain the data of 4 Periods, it covers the SourceForge status when it was first started, continued to exist, and when we did our analyses. Figure 3 shows that the degree distribution of the network model in each period (II to IV) follows a stretched exponential distribution [3], indicating that these network models belong to complex networks.

---

[3] A stretched exponential distribution (cumulative distribution) $P_C(x)$ is: $P_c(x) = exp\left[-\left(\frac{x}{x_0}\right)^c\right]$, where exponent $c$ is smaller than 1.

**Table 2. SourceForge Network Model Properties in Each Period**

| Period→ | 2000/3– 2000/12.2 | 2000/3 –2008.7.5 (I) | 2011/8/26 –2012.2.26 (II) | 2012/2/27 –2012.8.27 (III) | 2012/8/28 –2013/2.28 (IV) |
|---|---|---|---|---|---|
| Number of Network Nodes | # | _152402_ | 26734 | 27739 | 59129 |
| Number of Stopped- project Nodes | 335 | 28703 | 26734 | 27739 | 59129* |
| Exponent c of Stretched Exponential Distribution of Degree | # | # | $0.576/R^2=0.987$ | $0.4050/R^2=0.9287$ | $0.4258//R^2=0.908$ |
| Number of Connected Graph/Proportion of Number of Isolated Nodes in It | # | _85226_ (including 24146 connected graphs composed of stopped projects) | 20066/0.862 | 19221/0.848 | 34412/0.846 |
| Number of Isolated Nodes / Proportion in Network Nodes | # | _67626_ (including 20738 Stopped- project nodes) _/0.444_ | 17302/0.647 | 16298/0.588 | 29100/0.492 |
| Size of Maximum Connected Graph/ Proportion in Network Nodes | # | It can be found that the largest connected graph among the stopped projects contains 18 nodes | 1560/0.058 | 3233/0.117 | 15967/0.270 |

Note 1: The network nodes in Period I represent all kind of projects, so whose network model indicators have different meanings from those in other Periods and are indicated with _italics_. The network nodes in Period II-IV represent the stopped projects. Note 2: The network model in Period I is based on common principal developers, the property of the network model is same with the network model based mixed relations in Period I. The networks in Periods II-IV are based on mixed relations. Note 3: 59129* see Section 6.

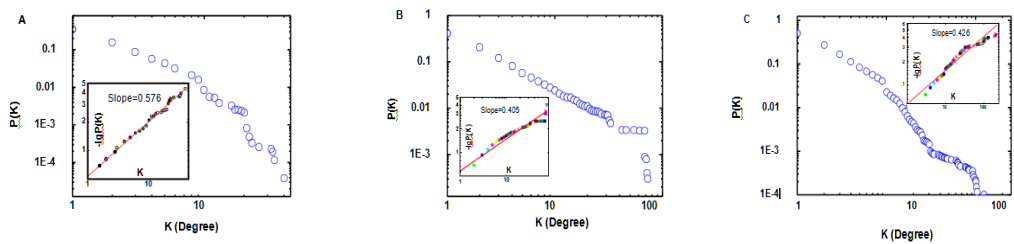

**Figure 3. Degree Distribution of Network Models in Periods II (A), III (B), and IV (C) Follows a Stretched Exponential Law**

Second, by replacing the "network language" in Table 2 with "cascade language", we can obtain the cascade size characteristics of SourceForge in Period II-IV (Table 3). We can also obtain the SourceForge cascade size characteristics in Period I from the network model of all types of projects of Period I. From Table 3, it can be observed that the cascade sizes in SourceForge exhibit the following characteristics: (1)

From the perspective of one-project cascades, first, in each Period, the proportion of one-project cascades in all cascades exceeds 0.846. In Period I, the number of cascades (stopped- project connected graphs) is 24146, in which the number of one-project cascade (isolated nodes) is 20738 (the ratio is 0.859), and the largest cascade size is just 18 (happened 1 time); Moreover, in the first year, i.e., March 2000 to December 2000 (included in Period I), the largest cascade size is just 2 (happened 19 times), the other 297 times are one-project cascades (the ratio is 0.940). We also know that in Periods II to IV, there are respectively 20,066, 19,221, & 34,412 occurrences of cascades, with 0.862, 0.848, and 0.846 of them being one-project cascades, which shows a slight decreasing trend. Second, there are respectively 26,734, 27,739, and 59,129 stopped projects in Periods II to IV, the proportion of the stopped projects included by one-project cascades in total stopped projects exceeds 0.492, but it decreases from 0.647 in Period II to 0.588 in Period III and then 0.492 in Period IV, indicating that the connections among projects are increasing more and more. (2) From the perspective of the largest cascades, in Periods II to IV, the largest cascade sizes (i.e., the number of stopped projects included in each largest connected graph) are 1,560, 3,233, and 15,967, respectively; they account for 0.058, 0.117 & 0.270 of total stopped projects in Periods II to IV, respectively, indicating that the sizes of largest cascades are relatively large and show a clear growth trend. From the above two points, it can be known that for each of Periods II to IV, there are many smallest one-project cascades (accounting for over 84.6% of all cascade events and involving over 49.2% of stopped projects), as well as one very big cascade with a large number of stopped projects (accounting for over 5.84% of stopped projects). (3) From the perspective of heterogeneity, in Periods I, the smallest cascade size is 1 and the largest cascade size is only 18, therefore, we can say that in Periods I, the cascade sizes in SourceForge are basically non-heterogeneous. In Periods II to IV, the heterogeneity indicators ( see Section 2) is 4, 4, and 5, respectively, so it can be said that the cascade size in Periods II to IV exhibit strong heterogeneity. (4) From the perspective of number of successful projects, there are almost no successful projects in the largest cascades in Period I, whereas the proportion of the successful projects in the largest cascades in Periods III and IV reaches 13.9% (448/3,233) and 44.6% (7,126/15,967), respectively, also displaying an clear increasing trend. The higher the proportion, the higher is the efficiency of the OSSOC.

**Table 3. Cascade Size Characteristics in SourceForge across Different Periods**

| Period→ | 2000/3–2000/12 | 2000/3-2008.7.5 (I) | 2011/8/26–2012.2.26 (II) | 2012/2/27–2012.8.27 (III) | 2012/8/28–2013/2.28 (IV) |
|---|---|---|---|---|---|
| Number of Stopped Projects | 335 | 28703 | 26734 | 27739 | 59129* |
| Number of Cascades (including One-Project Cascades) | 316 | 24146 | 20066 | 19221 | 34412 |
| Number of One-Project Cascades/ Proportion in Cascades/ Proportion in total Stopped Project | 297/0.940 | 20738/0.859/0.723 | 17302/0.862/0.647 | 16298/0.848/0.588 | 29100/0.846/0.492 |
| Size of Largest Cascade/Number of Times it occurs/ Proportion in total stopped projects | 2/19 | 18/1/# | 1560/1/0.058 | 3233/1/0.117 | 15967/17/0.270 |
| Number of successful projects in largest cascades | 0 | 0 | - | 448 | 7,126 |
| Exponent/ $X_{min}$/ Proportion (P) of Not Be Rejected of PL Distribution of Cascade Sizes | # | 4.4103/2/P=1 | 3.2624/1/P=1 | 3.1707/1/P=0.996 | 3.5592/2/P=0.994 |

Third, we further applied the Clauset method to fit a PL distribution to the cascade size data from each Period in SourceForge. After completing all necessary steps outlined in the Appendix "Procedures for Analyzing the Distributions of Datasets", we find that for the cascade sizes in each Period, the parameter $X_{min}$ is 2, 1, 1, 2, respectively, and the parameter α is 4.4103, 3.2624, 3.1707, and 3.5592, respectively, as shown in Figure 4. Out of 2,500 KS tests performed on the cascade sizes for each Period, 2,500, 2,500, 2,488, and 2,484 tests failed to reject the null hypothesis, respectively. This indicates that the PL distribution provides a good fit to the dataset for each Period. For all cascade size datasets, we also conducted goodness-of-fit tests against other relevant distributions and find that they do not fit the data well. Therefore, we conclude that the PL distribution is indeed a good fit for the cascade size datasets.

Based on the fitting results above (Figure 4), it can be seen that although the PL distribution fits the cascade sizes well for each Period, all the obtained PL exponents have absolute values greater than 3. Consequently, the cascade sizes in each Period do not exhibit scale-free characteristics. This conclusion can be explained from the heterogeneity and SOPLD. In Period I, the PL fits doesn't not include the one-project cascades and the cascade sizes are non-heterogeneous, thus the cascade sizes do not have the scale-free characteristic; In Periods II to IV, the PL fits do not include the largest cascade and in Period IV it also doesn't not include the one-project cascades, with their SOPLD indicators being 2/4, 2/4 and 2/5 respectively. Although in Periods II to IV, the cascade sizes all are heterogeneous, their SOPLD indicators do not exceed 50% and each SOPLD doesn't include the largest cascade or one-project cascades, so the cascade sizes in Periods II to IV of SourceForge also do not have the scale-free characteristic.

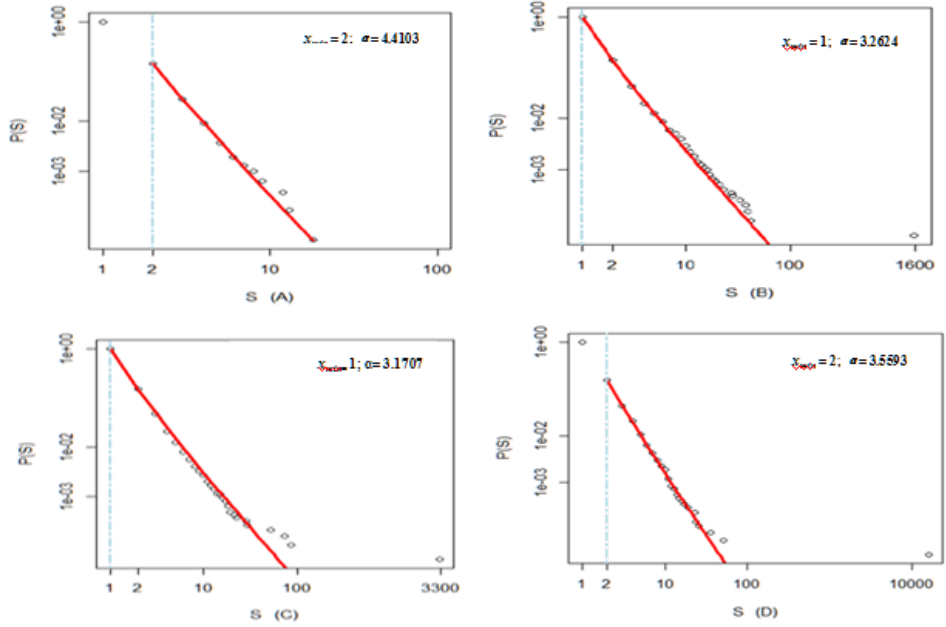

**Figure 4. The Cumulative Distribution of Cascade Sizes P(S) and Their Maximum Likelihood PL Fits (to the line); (A) Period I; (B) Period II; (C) Period III; (D) Period IV.**

## 6. Discussion

**Limitations**

The first limitation comes from our data extraction: Using the software, LocoySpider, to extract the data from the SourceForge website is limited by some objective conditions, so our data for analyzing the cascade sizes are incomplete and not totally accurate. For instance: (1) When we did the second round of data extraction on Feb. 28, 2014, there were roughly 200,000 stopped projects, but we could only use the 114,016 stopped projects that had the information for analyzing the cascade sizes; this resulted in a significant reduction in the number of cascades between the one-project cascade and the largest cascade. (2) The collected stopped projects in the last Period (i.e., Period IV) have a time interval of one year or more between their last update date and our data extraction date, and for the stopped projects in the earlier Periods, the time intervals are 1.5 years or more. Thus the collected stopped projects in the last Period would include some number of projects that actually have not stopped (see Section 3), so the amount of stopped projects in Period IV (59129* in Table 2) is more than the actual amount; (3) The data on the website are varying; therefore we can't repeatedly extract the same data for a same Period at different data extraction dates, which means that we can't later check the accuracy of our collected data for each Period.

The second limitation comes from methods. The limitation in data extraction further leads to limitation in methods. For instance, we use the "longest duration of a cascade" as the criterion for dividing Periods (II to IV), which is determined by the "snapshot method" based on the extracted data. Due to the limitation in data extraction, it is difficult to ensure that each Period divided by this criterion is in the same state.

**Issues related to SOC theory**

Since SourceForge can be metaphorically described as a "sandpile" of projects and at first glance it also possesses the key elements of a BSOS (Section 2), we also fitted the distributions of cascade sizes in its deferent Periods, it seems that further analysis should be conducted to determine whether it has evolved into the critical state of SOC or is evolving towards this critical state. However, we do not do so for the following reasons:

On one hand, Bak determined whether a system had reached the SOC state by examining whether its sandpile model, maintained at a constant average slope, exhibited scale-free properties in its cascade sizes. However, as our study focuses on the real world rather than a laboratory problem, even if SourceForge were truly a BSOS, we cannot maintain a constant average slope of the SourceForge "sandpile" in the real world, as we would in a laboratory, to obtain data on its multiple cascade sizes.

The actual approach we have adopted is as follows: First, we infer logically that the SourceForge state (i.e., the average slope of its "sandpile") only changes after the longest cascade ends, so we assume that the SourceForge "sandpile" maintains a constant average slope (i.e., the system's state remains unchanged) throughout the duration of the longest cascade. Then, using this duration of the longest cascade as a criterion, we divide the SourceForge evolution process (except Period I) into several Periods (each Period being in the same state), thereby obtaining the data of the cascade sizes for Periods II to IV. It is clear that the above assumption: "SourceForge "sandpile" maintains a constant average slope throughout the duration of the longest cascade" has not been verified by us and is difficult to verify. Moreover, given the limitations of data and methods, our specific estimate of the "longest duration of a cascade" also have a deviation from the actual value.

On the other hand, from a fundamental logic perspective, there is no consensus on whether SourceForge as an OSSOC is truly a BSOS that can be analyzed using SOC theory. This is because, first, it is difficult to strictly verify that the SourceForge "sandpile" is being slowly driven (see Section 2). Second, during its peak period, the SourceForge cascades sizes are either extremely large or extremely small, with few of intermediate sizes. Additionally, the distributions of cascade sizes do not exhibit the scale-free characteristics and do not show a clear evolutionary trend towards scale-free characteristics. Third, we further recognize that from Checkland's classification of systems (Checkland, 1981), most of the systems currently successfully applying SOC belong to natural systems or man-made (designed) systems. For example, physical systems related to earthquakes and ecosystems related to forest fires or species extinction events are all natural systems, whose complex behaviors are always created by a long process of evolution (Bak 1996, p. 31), while sandpiles in laboratories belong to man-made systems. SourceForge is a human activity system constrained by its software development digital platform (man-made system), which differs from natural or man-made systems as its elements are people with autonomous goal-directed behavior. Fourth, although SOC theory successfully explains the phenomenon of phantom traffic jams, the constraints imposed by a certain road segment (a human-made system on which traffic flows operate) on its traffic-flow participants are stronger than those imposed by the SourceForge platform on its participants. Consequently, the characteristics of SourceForge's human activity system are more pronounced. In other words, both SourceForge and the traffic flow on a certain road segment belong to constrained human activity systems, but SourceForge participants operate under far fewer restrictions than their counterparts in the traffic flow. Therefore, SourceForge participants exhibit stronger agency—they can enter and exit the OSSOC more freely—which may lead to its abrupt decline or even disappearance, whereas the traffic flow on a certain road segment can persist for a longer duration.

Based on the above analysis, we suggest that SOC theory may not be applicable for analyzing the evolving process of an OSSOC. However, due to limitations in our data extraction, the distributions of cascade sizes in SourceForge fitted to these data may be inaccurate. Thus, we cannot validate the above perspective and can only raise the question: Is SOC theory applicable to OSSOCs?

## Appendix: Procedures for Analyzing the Distributions of Datasets

The fitting of PL distributions to a dataset includes following steps (Lai, 2016).

**Observing Data**: This step is to observe whether the shape of the histogram follows some kind of PL distribution.

**Estimating Parameters**: This step is to estimate the two important parameters $\alpha$ and $x_{\min}$ used in the following equation:

$$P(x) = (x/x_{min})^{(-\alpha+1)} \qquad (2)$$

where $x$ is the collected data, $P(x)$ is the cumulative distribution function of $x$. The starting point of a PL fit is $x_{min}$: (1) we pick the minimum of the Kolmogorov-Smirnoff (KS) statistic to find the $x_{min}$ value; and (2) we use MLE (maximum likelihood estimation) to compute a plausible value for $\alpha$.

**Testing for Goodness of Fit**: In 2500 KS tests, if the results show that there are less than 10% of the tests that reject the null hypothesis, we can conclude that the PL distribution is a good fit to the observed data (null hypothesis).

**Testing for other Possible Distributions:** Although a PL distribution may show a good fit to the dataset, this doesn't mean that the PL is the best fit. Other distributions may be just as good or even have better fits. Thus, we need to fit each distribution to the data and perform a goodness of fit test to see if the fits of other kinds of distributions are better than a PL distribution. If these other distributions are not better fits, then the PL may be considered as having the best fit to the data (Lai 2016).

**Acknowledgments**

Jianmei Yang acknowledges the support from NNSF of China grant #71273093. Thanks to Liu Xiao's, Sha Sha's and Weicong Xie help in data extraction and processing.